\documentclass[journal]{igtesymp}
\ifCLASSINFOpdf
\else
\fi
\usepackage{amsmath,mathrsfs}
\usepackage{amsfonts,amssymb}

\usepackage{siunitx}

\usepackage{mhchem}

\newcommand{\figref}[1]{Fig.~\ref{#1}}

\renewcommand{\eqref}[1]{Eq.~$($\ref{#1}$)$}
\newcommand{\eqsref}[2]{Eqs.~$($\ref{#1}$)$-$($\ref{#2}$)$}

\usepackage{graphicx}

\begin{document}
\newcommand*{\E}[1]{\mathop{}\!\times 10^{#1}}

\def\diff{\mathrm{d}}

\def\mag{\mathbf{m}}
\def\eau{\mathbf{u}}
\def\pos{\mathbf{r}}
\def\hh{\mathbf{h}}
\def\H{\mathbf{H}}
\def\M{\mathbf{M}}
\def\J{\mathbf{J}}
\def\B{\mathbf{B}}

\def\Hn{H_\mathrm{n}}
\def\Hp{H_\mathrm{p}}
\def\Hn{H_\mathrm{c}}

\def\Hsw{H_\mathrm{sw}}

\def\Ae{A_\mathrm{ex}}
\def\Ku{K_{\mathrm{u}1}}
\def\sigdw{\sigma_\mathrm{dw}}
\def\lex{l_\mathrm{ex}}

\def\delFdelm{\frac{\delta \mathcal{F}}{\delta \mathbf{m}}}
%
\title{Micromagnetics and Multiscale Hysteresis Simulations of Permanent Magnets}
%
%
%

\author{Yangyiwei Yang\IEEEauthorrefmark{1}, Patrick Kühn\IEEEauthorrefmark{1}, Mozhdeh Fathidoost\IEEEauthorrefmark{1}, and Bai-Xiang Xu\IEEEauthorrefmark{1} 
 \\ \vspace{0.3cm} \normalsize{
\IEEEauthorblockA{\IEEEauthorrefmark{1}Mechanics of Functional Materials Division, Institute of Materials Science, Technische Universität Darmstadt, 64287 Darmstadt, Germany\\
 E-mail: xu@mfm.tu-darmstadt.de,  yangyiwei.yang@mfm.tu-darmstadt.de}}
}

%
%


\IEEEaftertitletext{\vspace{-1cm}\noindent\begin{abstract} 
Confronting the unveiled sophisticated multiscale structural and physical characteristics of hysteresis simulation of permanent magnets, notably samarium-cobalt (Sm-Co) alloy, a novel scheme is proposed linking physics-based micromagnetics on the nanostructure level and magnetostatic homogenization on the mesoscale polycrystal level. Thereby the micromagnetics-informed surrogate hysteron is the key to bridge the scales of nanostructure and polycrystal structure. This hysteron can readily emulate the local magnetization reversal with the nanoscale mechanisms considered, such as nucleation of domains, and domain wall migration and pinning. The overall hysteresis, based on a sintered Sm-Co polycrystal, considering both mesoscale and nanoscale characteristics, is simulated and discussed. 

\end{abstract}
\noindent\vspace{\baselineskip}}

\maketitle
\thispagestyle{empty}\pagestyle{empty}

%
\IEEEpeerreviewmaketitle

\section{Motivations}
\begin{figure}[!t]
\centerline{\includegraphics[width=8cm]{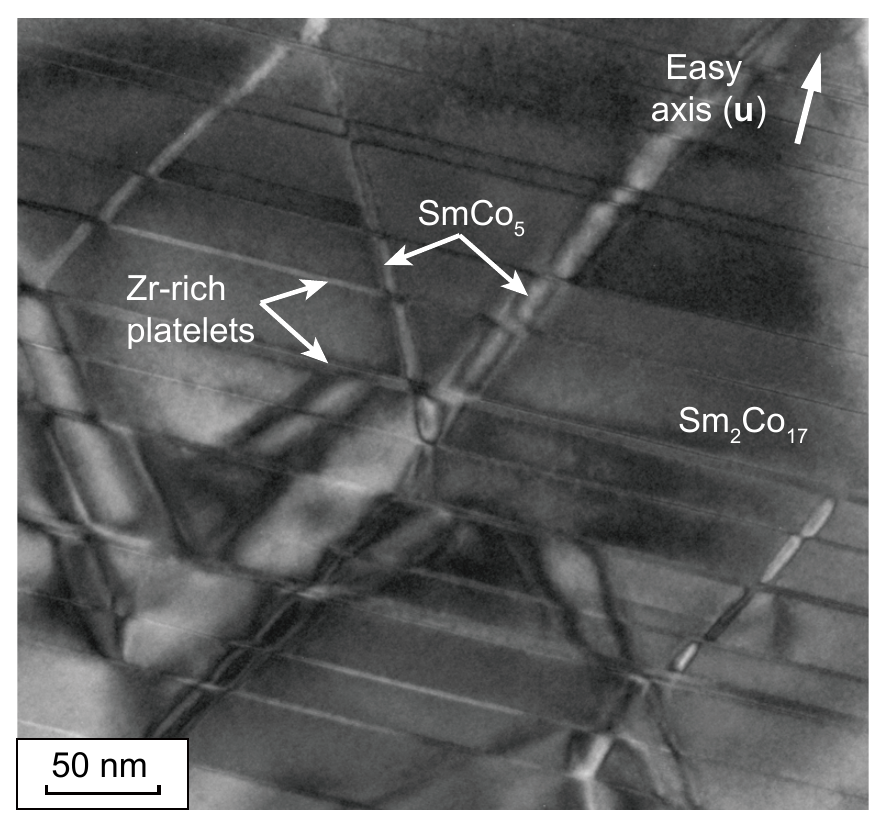}}
\caption{Bright-field TEM image of the Sm-Co magnets, where the nanostructure containing \ce{SmCo5}, \ce{Sm2Co17} and Zr-rich phases. The image is reprinted from \cite{duerrschnabel_atomic_2017} under the terms of the Creative Commons CC-BY license.}
\label{fig:tem}
\end{figure}

Due to its superior corrosion resistance under complex chemical environment and outstanding stability at high temperature, samarium-cobalt (Sm-Co) based magnets has promised industries feasibility in various applications, such as high-performance electric motors for automotive and aeronautic application. However, former researches have unveiled the microstructure of such permanent magnets in a sophisticated multiscale fashion. As shown in \figref{fig:tem}, it has been investigated that commercial Sm-Co magnets present a three-phase composite nanostructure ($<1$ \si{\micro m}). This nanostructure can be described as the cellular Sm$_2$Co$_{17}$ phase surrounded by a coherent stripe-shaped SmCo$_5$ phase. This is further subdivided by the Zr-rich platelet-shaped phases (hereinafter referred to as Z-platelets) which develop perpendicular to the \textit{c}-axis \cite{duerrschnabel_atomic_2017, katter1996new, song_cell-boundary-structure_2020}.
The final nanostructure depends on the chemical composition and thermal treatments \cite{wang_overview_2021}. Notably, Zhou et al. \cite{zhou_revisiting_2021} reported increases in thickness of Z-platelets from 2.4 \si{\nano\meter} up to 28.8 \si{\nano\meter} with only an increase in annealing time. Meanwhile, as one of the commercial permanent magnets that is  manufactured by sintering, polycrystalline structure on the mesoscale (1-100 \si{\micro m}) is also observed and examined in Sm-Co alloy \cite{giro2022}. 

Most of the primary mechanisms contributing to the magnetic behavior (i.e., nucleation of the reversed domain, and migration/pinning of the domain wall) occur on the nanoscale with characteristic length around 1 \si{nm}, while receiving effects from grain orientation and local thermal history, hysteresis behavior varies locally on the level of the polycrystal. This stresses the importance of scale-bridging on accurate modeling and simulation of the hysteresis behavior. Although there are well-established models for individual scales, scale-bridging strategy is intricate and essential. It is worth noting that the strategy bridging the atomic and nanoscale micromagnetic combining first-principles calculations, atomistic spin model simulations, and micromagnetic simulations has been investigated and discussed \cite{tuprints22038}. Nevertheless, the strategy bridging the nanoscale and mesoscale is still missing. On the other hand, macroscopic hysteresis behavior can be directly modeled using proposed phenomenological models, such as the Preisach's \cite{preisach1935magnetische}, T\^{a}k\^{a}cs's \cite{takacs2001phenomenological} and Jiles-Atherton's model \cite{jiles_theory_1984, zirka_physical_2012}. These models, however, fail to deliver the physics information on individual scales and cannot be used in the sense of tailoring hysteresis of permanent magnets.

At the end of the day, modeling and simulating the hysteresis of the permanent magnets demands a multiscale scenario bridging mesoscale phenomena and nanoscale mechanisms, which becomes the objective of this work. We propose a novel multiscale scheme for simulating the polycrystalline permanent magnets' hysteresis combining the merits of both micromagnetics and computational magnetostatic homogenization. This scheme is also extendable to simulate the part-level hysteresis, and is capable of integrating machine-learning-based data-driven methods. 
It is hoped that the present work can serve a new viewpoint/methodology in the field of electromagnetic engineering in the hysteresis behavior of magnetic materials and components and provide a computational toolkit that is practicable and physics-rooted. 


\section{Models and methods}
\begin{figure}[!th]
\centerline{\includegraphics[width=8cm]{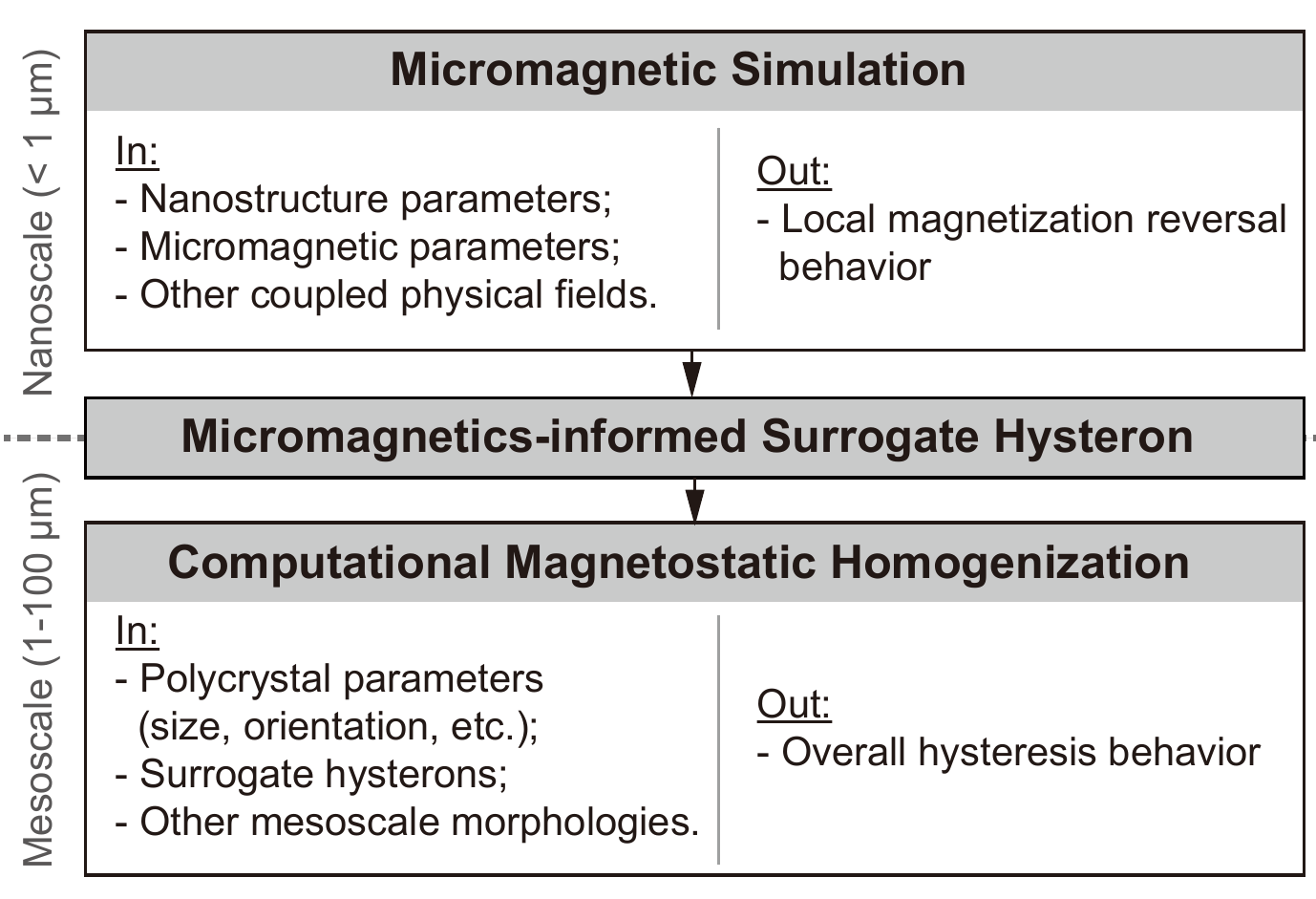}}
\caption{Workflow of proposed multiscale hysteresis simulation scheme.}
\label{fig1}
\end{figure}

\figref{fig1} presents the workflow of proposed multiscale hysteresis simulation for the polycrystalline permanent magnets. We start with performing a series of micromagnetic simulations on distinct parameterized nanostructures. Micromagnetics has a sound physics foundation and thus is suitable for investigating local magnetization switching mechanisms. To have excellent computational cost-efficiency with fine spatial discretization for resolving physical processes (e.g. domain nucleation, and domain wall migration and pinning), micromagnetic simulations are implemented and performed by the finite difference method (FDM). Next, the surrogate hysteresis unit (or `hysteron', adopted from its pseudo-particle behavior) parameterized by the results of micromagnetics is employed to replace micromagnetic calculations on the mesoscale polycrystalline structures. This hysteron should preserve the physical characteristics unveiled by the micromagnetics, such as the local magnetic coercivity and magnetization rotation. Finally, the hysteresis behavior of the polycrystalline structure with micromagnetics-informed hysterons is evaluated by computational magnetostatic homogenization. Due to the need for geometrically-complicated polycrystalline structures without compromising in numerical accuracy, magnetostatic homogenizations are implemented and performed by the finite element method (FEM).

It is worth noting that the magnetization reversal on each level of the proposed scheme is modeled and simulated under the equilibrium conditions, which is sufficient for evaluating the material- and structural-based hysteresis behavior without loss. It also helps reduce the 3D magnetization dynamics (a.k.a. Landau-Lifshitz-Gilbert dynamics) to 2D rotational one (a.k.a. Stoner-Wohlfarth reversal, where magnetization is also in the plane defined by the magnetic field $\H$ and the easy axis $\eau$).



\subsection{Micromagnetics}
Below the Curie temperature, the magnetization of most of magnetic materials saturates with constant magnitude ($M_\mathrm{sat}$). Therefore in the micromagnetics, it is important to have the normalized magnetization vector which is position dependent, i.e., $\mag(\pos)$. This vector field can be physically interpreted as the mean field of the local atomic magnetic moments, but yet sufficiently small in scale to resolve the magnetization transition across the domain wall. In this regard, We consider the free energy functional of a micromagnetic system (with a volume $V$) as the functional of $\mag(\pos)$, i.e.,
\begin{equation}
    \mathcal{F}=\int_V \left[f_\mathrm{ex}+f_\mathrm{ani}+f_\mathrm{ms}+f_\mathrm{zm}+f_\mathrm{cp}\right]\diff V.
    \label{eq:F}
\end{equation}

Hereby $f_\mathrm{ex}$ is the exchange contribution, recapitulating the parallel-aligning tendency among neighboring magnetic moments due to the Heisenberg exchange interaction. In that sense, this term acts as a thermodynamic penalty to the system at the domain wall, and provides the local driving force to the domain wall migration. By defining a positive exchange parameter $A_\mathrm{ex}$, this term is formulated as
\begin{equation}
    f_\mathrm{ex} = A_\mathrm{ex} \|\nabla\mathbf{m}\|^2.
\end{equation}

The term $f_\mathrm{ani}$ represents the contribution due to the magneto-crystalline anisotropy. It provides the energetically preferred orientation to local magnetizations, which is related to the defined easy axis $\eau$ (normally the principal axis) of the crystal lattice. For most of the permanent magnets, whose lattice possesses the uniaxial symmetry, this term is formulated as
\begin{equation}
    f_\mathrm{ani} = -\sum_i  K_{\mathrm{u}i}\left(\eau\cdot\mag\right)^{2i}.
\end{equation}
It is worth noting that most of the investigations only employ the lowest order ($i=1$) with the characteristic parameter $\Ku$. It can be shown that the parameters $\Ae$ and $\Ku$ are related to the domain wall energy $\sigdw$ and the exchange length $\lex$ of the materials at the equilibrium as $\sigdw=4\sqrt{\Ae\Ku}$ and $\lex=\pi\sqrt{\Ae/\Ku}$, using the SI units.

Besides $f_\mathrm{ex}$ and $f_\mathrm{ani}$ which are the material-dependent, the terms $f_\mathrm{ms}$, $f_\mathrm{zm}$ and  $f_\mathrm{cp}$ provide the contributions due to the interaction among magnetization and distinct intrinsic/extrinsic fields and thereby change the local thermodynamic stability. The magnetostatic term $f_\mathrm{ms}$ counts the energy of each local magnetization (or a magnetic moment) under the demagnetizing field created by the surrounding magnetization (or by all the other magnetic moments). It is generally formulated as
\begin{equation}
    f_\mathrm{ms} = -\frac{1}{2}\mu_0M_\mathrm{sat}\mag\cdot\H_\mathrm{dm},
\end{equation}
where the calculation of the demagnetizing field $\H_\mathrm{dm}$ highly depends on the choice of boundary condition (BC), and $M_\mathrm{sat}$ is the saturated magnetization of the material. Similarly, the Zeeman and multiphysics-coupling terms $f_\mathrm{zm}$ and $f_\mathrm{cp}$, correspondingly counting the energy of each local magnetization under an extrinsic magnetic field $\H$ and an effective field induced by the coupled physical effects \textbf{$\H_\mathrm{eff}$}, are formulated as
\begin{align}
    f_\mathrm{zm} &= -\mu_0M_\mathrm{sat}\mag\cdot\H,\\
    f_\mathrm{cp} &= -\mu_0M_\mathrm{sat}\mag\cdot\H_\mathrm{eff}.
\end{align}
The exact formulation $\H_\mathrm{eff}$ depends on the choice of coupled physical effects, such as the magnetostriction \cite{kronmuller2003micromagnetism}, the thermal fluctuation \cite{brown1963thermal}, and spin-current interactions \cite{sun2000}.

Generally, $\H$ is implemented as a controllable quantity for the investigator to emulate the magnetic loading/unloading as in the experiments, i.e., an applied magnetic field $\H_\mathrm{ext}$, and the micromagnetic system should find its equilibrium. This is mathematically determined by
\begin{equation}
\begin{split}
&\mathbf{m} \times \delFdelm + \alpha\mathbf{m} \times\left(\mathbf{m} \times \delFdelm\right) = \mathbf{0} \\
&\text{subject to}\quad\|\mathbf{m}\|=1,\label{eq:gov_mm}
\end{split}
\end{equation}
and the solved $\mathbf{m}(\mathbf{r})$ will be the thermodynamically-preferred local magnetization of the system. $\alpha$ is the damping coefficient. FDM is considered to be relatively fast, cost-efficient, and readily for GPU-parallel implementation, but incompatible in adequately discretizing complicated (curved) geometries and efficiently dealing with sparse problems \cite{leliaert2018fast}. In this work, the FDM-based steepest conjugate gradient (SCG) method is employed. The core of this method is to calculate the next magnetization $\mathbf{m}_{n+1}$ on the $n$-th iteration step by a descending direction $\mathbb{H}_n$ by 
\begin{equation}
    \mathbf{m}_{n+1}=\mathbf{m}_n-\tau_n\mathbb{H}_n,
\end{equation}
with
\begin{equation*}
\mathbb{H}_n=\mag_n^{*}\times\nabla_\mathbf{m}\hat{\mathcal{F}}(\mag_n,\H)\times\mathbf{m}_n.
\end{equation*}
Here $\mag_n^{*}=\mag^n$ will derive $\mathbb{H}_n$ as the steepest searching direction, and $\mag_n^{*}=(\mag_n+\mag_{n+1})/2$ as the curvilinear searching direction on the sphere \cite{goldfarb2009curvilinear}. $\hat{\mathcal{F}}(\mag_n,\H)$ is the free energy on the discretized domain. The step length $\tau_n$ is initialized by an inexact line search and subsequentially obtained by the Barzilai-Borwein rule. This method has been implemented by  FDM in the package \texttt{MuMax$^3$} \cite{Vansteenkiste2014} with details elaborated in Ref. \cite{exl2014labonte}.

\subsection{Micromagnetics-informed Surrogate Hysteron}
To correctly simulate the local magnetization reversal and associated domain wall migration, sufficiently fine space discretization (mesh) is required to resolve the reversed nucleus formed near grain edges/corners where the demagnetizing field is high \cite{Yi2016a}, and the transition profile of magnetization across the domain wall with its thickness characterized by $\lex$. As most permanent magnets with $\lex$ in the range of several nanometers, e.g., $\lex\sim2~\si{nm}$ for Nd-Fe-B and Sm-Co magnets and $\lex\sim 50 ~\si{nm}$ for electrical steels ($4.6\%$ Si), it is impractical in both numerical and computational senses for direct micromagnetic calculations on the mesoscale structure of such materials, where the spatial distribution of grains with distinct sizes and orientations (easy axes) are believed to have the significant influences as well.
In this regard, a surrogate model is required to equivalently replace the direct micromagnetic calculation on every subdomain for the hysteresis simulation on the polycrystal level. Such a surrogate model
\begin{enumerate}
    \item can efficiently describe the local magnetization reversal as an isolated unit but also as the representative component (subsystem) of the polycrystal system. Such local reversal should be also dependent on the given orientation of the subdomain;
    \item can preserve the important characteristics of the local magnetization reversal by the micromagnetics, e.g., the local magnetic coercivity $H_\mathrm{c}$ where the magnetization of the chosen subsystem cannot withstand the applied field and gets reversed;    
\end{enumerate}

\begin{figure}[!t]
\centerline{\includegraphics[width=8cm]{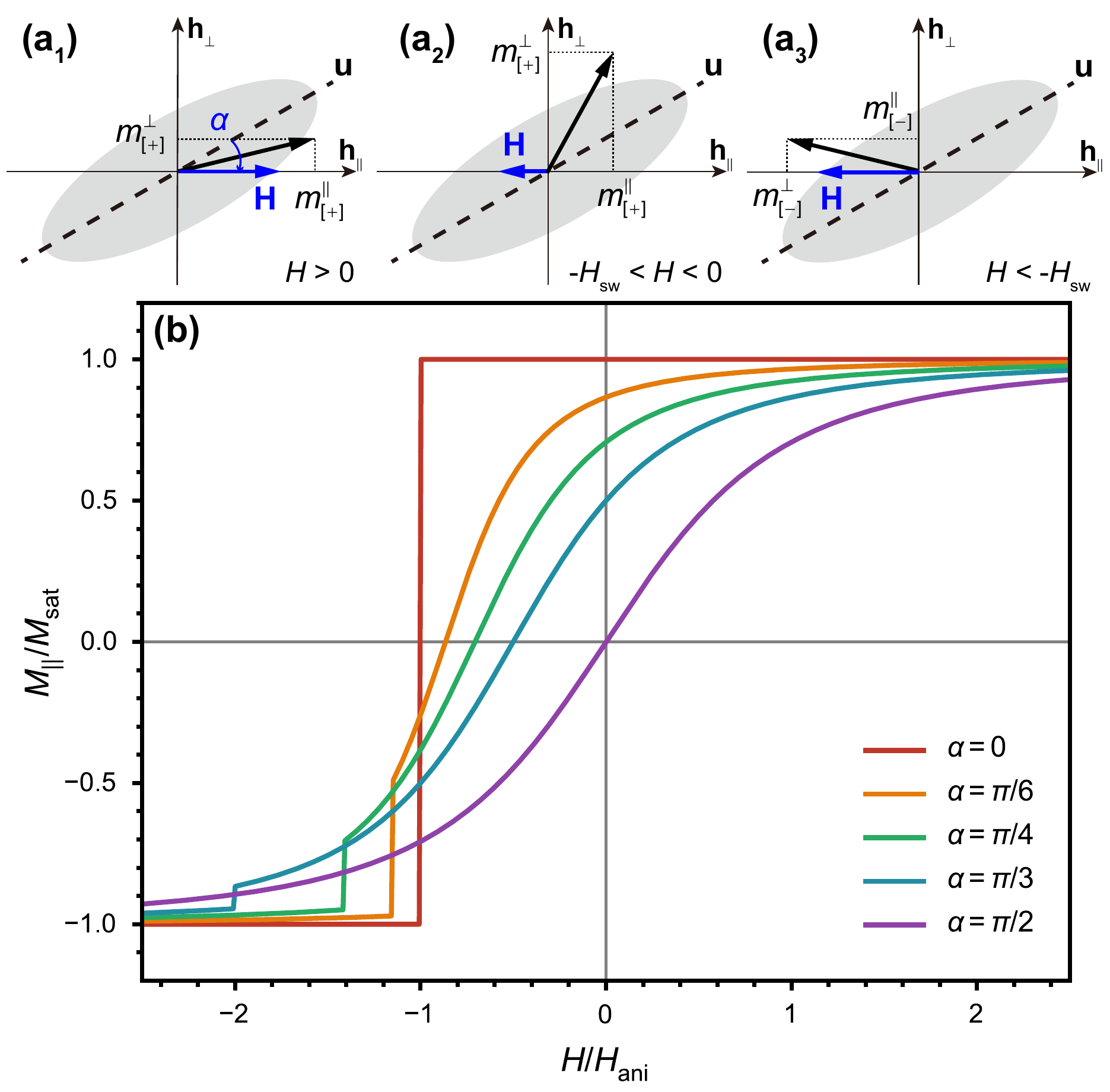}}
\caption{Pesudo-particles schematic of the vector hysteron with easy axis $\mathbf{u}$, which present the equilibrium magnetization at varying field: (a$_1$) $H>0$; (a$_2$) $-\Hsw<H<0$; (a$_3$) $H<-\Hsw.$ $m_{[\pm]}^{\|}$ and $m_{[\pm]}^{\perp}$ are respectively the longitudinal and transverse magnetization components with $[+]$ for the upper branch and $[-]$ for the lower branch. $\H=H 
\hh_{\|}$. (b) Longitudinal half-cycle hysteresis of parameterized vector hysteron with different easy axis orientations. }
\label{fig2}
\end{figure}

In this work, we use the vector hysteron as the surrogate model of the micromagnetic simulations, which can well describe the magnetization reversal of the ferromagnetic domain at equilibrium. Each vector hysteron can be regarded as an independent Stoner-Wohlfarth pseudo-particle as its magnetization can only rotate freely in the plane defined by local magnetic field $\H$ and the easy axis $\eau$ (if $\H\|\eau$, then the hard axis perpendicular to $\eau$ should be provided instead), as shown schematically in inset of \figref{fig2}. A hysteron inside a system (as an assembly of hysterons) can only affect the neighboring one via magnetostatic interactions, in other words, by affecting the local magnetic field.

This vector hysteron consists of two major parameters: the local switching field $\Hsw$ and the orientation angle $\alpha$. Defining the local coordinates by the defined positive direction of the applied magnetic field, i.e., $\H=H\mathbf{h}_\|$, the longitudinal magnetization of a single demagnetizing process (simplified as $H$ reversely increasing) is analytically formulated as
\begin{equation}
    \begin{split}
\mathbf{m}(H)=\left\{
\begin{aligned}
&m^{||}_{[-]}\mathbf{h}_\mathrm{||}+m^{\perp}_{[-]}\mathbf{h}_\mathrm{\perp}\quad& H<-H_\mathrm{sw}\\
&m^{||}_{[+]}\mathbf{h}_\mathrm{||}+m^{\perp}_{[+]}\mathbf{h}_\mathrm{\perp}\quad& \text{otherwise}
\end{aligned}
\right.
\end{split}\label{eq:VH}
\end{equation}
with
\begin{align*}
    m^{||}_{[\pm]}=\frac{H \pm H_\mathrm{sw}\cos \alpha}{\sqrt{H^2 \pm 2 H H_\mathrm{sw} \cos \alpha+H_\mathrm{sw}^2}}
\end{align*}
with the plus sign for the upper branch and the minus sign for the lower branch \cite{petrila2011analytical}. The transverse magnetization is the calculated$ m^{\perp}_{[\pm]}=\sqrt{1-(m^{||}_{[\pm]})^2}$ accordingly. The reversal of magnetization vector represented by \eqref{eq:VH} with prescribed $\alpha$ is schematically presented in \figref{fig2}a. The switching field $\Hsw$ should be intrinsically structure- and orientation-dependent and should be examined on varying nanostructures. We take the following $\Hsw$ to separate the dependence from a set of nanostructure parameters $\{s\}$ and the orientation $\alpha$
\begin{equation}
    \Hsw(\alpha, \{s\}) = \frac{H_\mathrm{ani}(\{s\})}{\cos\alpha} + H_\mathrm{l}(\alpha, \{s\}).
\end{equation}
where $H_\mathrm{ani}(\{s\})$ is physcially regarded as anisotropic field of a nanostructure. For homogeneous ferromagnetic materials, $H_\mathrm{ani}=\frac{2K_1}{\mu_0M_\mathrm{sat}}$. The longitudinal shifting field $H_\mathrm{l}(\alpha, \{s\})$ is to recapitulate the associate effects that might influence the magnetization reversal (like the pinning effects). Taking $H_\mathrm{l}=0$, the half-cycle hysteresis curves presented by \eqref{eq:VH} with varying $\alpha$ are illustrated in \figref{fig2}b, where we can observe the cut-off switching of the magnetization once reaching $\Hsw$ \cite{petrila2011analytical}. In this work, $\Hsw(\alpha)$ is the major parameter that is informed by micromagnetic simulations. It is expected to optimize the formulation of $\Hsw$ and further investigate its structural dependence adopting the data-driven scenario in our works in the near future.

\begin{figure*}[!ht]
\centerline{\includegraphics[width=18cm]{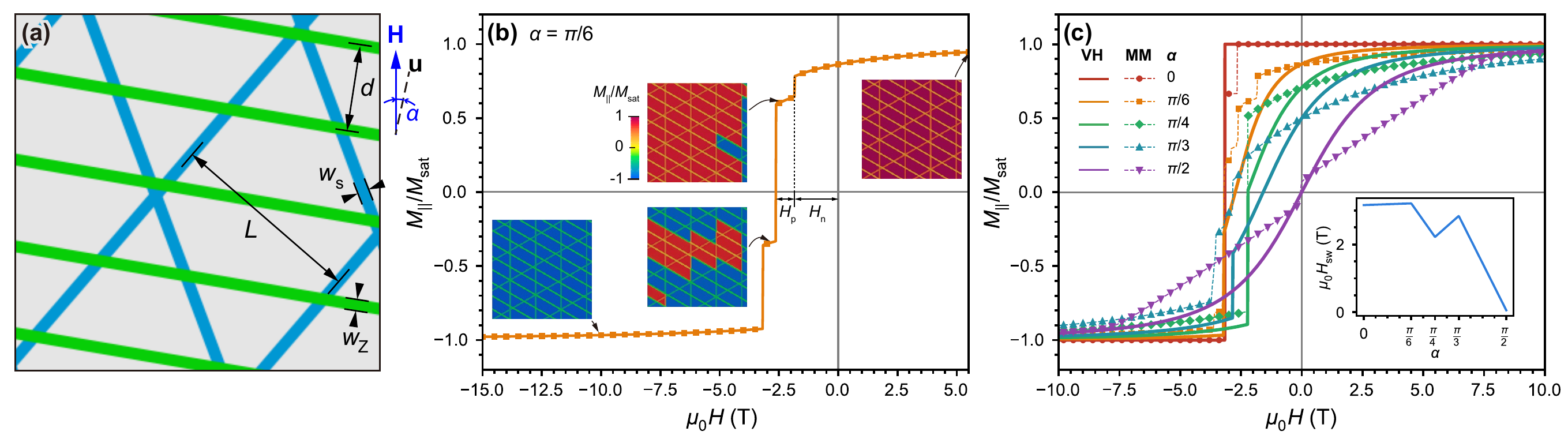}}
\caption{(a) Parameterized nanostructure of Sm-Co magnets. The positive direction of the magnetic field is also denoted. (b) The micromagnetic simulated half-cycle hysteresis of the single demagnetizing process of Sm-Co nanostructure with corresponding domain configuration denoted. (c) Average half-cycle hysteresis of five demagnetizing processes by micromagnetic simulation (MM) for each orientation angle ($\alpha$), which is compared with parameterized vector hysteron (VH). Inset: orientation dependence of the switching field ($\Hsw$).}
\label{fig3}
\end{figure*}

\subsection{Computational Magnetostatic Homogenization}
In this work, the overall hysteresis behavior of the polycrystal is examined by performing the computational magnetostatic homogenization. The governing equation for magnetostatics is derived by eliminating the time derivative terms in Maxwell's equations as
\begin{align}
\nabla\times\H&=\J,\label{eq:amp}\\
\nabla\cdot\B&=0, \label{eq:gauss}
\end{align}
where $\B$ is the magnetic flux density, and $\mathbf{J}$ is the current density. In this work we consider the case when no current density appears in the system ($\J=\mathbf{0}$), which means $\H$ is a curl-free vector field, i.e., $\nabla \times \mathbf{H}=\mathbf{0}$. In that sense, we can calculate the magnetic field by using merely the magnetic scalar potential $\Phi$, i.e., $\H=-\nabla\Phi$, and rewrite the Gaussian law  in \eqref{eq:gauss} in the Laplace form to be the governing equation of the system
\begin{equation}
    \Delta\Phi=0.\label{eq:gov}
\end{equation}
Omitting Ampère's law in \eqref{eq:amp} by assuming no space current density also disregards the effects of eddy current. The homogenization scheme considering the eddy current losses is in development and will be discussed in upcoming works.

The magnetostatic homogenization problem can then be defined as
\begin{equation}
    \left\{
\begin{aligned}
\mathbf{B}=&\mu_{0}\left[\mathbf{H} +  \mathbf{M}(\mathbf{H})\right]\quad&{\text{on microscale}},\\
\langle\mathbf{B}\rangle =&\mu_{0}\left[\langle\mathbf{H} \rangle + \langle\mathbf{M} \rangle\right]\quad&\text{on macroscale},
\end{aligned}\right.
\end{equation}
where $\langle\cdot\rangle=\int_{V}(\cdot) \mathrm{d} V / V$, and $V$ here is the volume of the simulation domain. By using the micromagnetics-informed surrogate hysteron, the local magnetization $\mathbf{M}(\mathbf{H})$ not only readily recapitulates the magnetization reversal determined by the nanostructure and micromagnetic contributions, but also reflects the orientation-dependences on the polycrystal level. $\langle\mathbf{H} \rangle$ is provided by linear BC of scalar potential that satisfies the Hill-Mandel condition $\langle\mathbf{B}\cdot\mathbf{H}\rangle=\langle\mathbf{B}\rangle\cdot\langle\mathbf{H}\rangle$, i.e.,
\begin{equation}
\begin{split}
\Phi|_{\partial V}=-\H_\mathrm{ext}\cdot\pos|_{\partial V},\label{eq:lbc}
\end{split}
\end{equation}
where $\H_\mathrm{ext}$ is constantly prescribed, and $\pos$ is the coordinates. \eqsref{eq:gov}{eq:lbc} are implemented by FEM in the package \texttt{NIsoS} developed by authors based on \texttt{MOOSE} framework \cite{tonks2012object, permann2020moose}. 4-node tetrahedron Lagrangian elements are chosen to mesh the geometry. A transient solver with preconditioned Jacobian-Free Newton-Krylov method (PJFNK) and backward Euler algorithm is employed. Randomly seeded polycrystalline structures as well as corresponding meshes are created using the open-source package \texttt{Neper} \cite{quey2011large, quey:hal-01626440, quey:hal-01850591}.  

\section{Preliminary Results and Discussion}

\subsection{Hysteresis of Sm-Co Nanostructure and its Orientation Dependence}
Following Katter et al., \cite{katter1996new} a parameterized nanostructure for Sm-Co is employed. The structure parameters of interest are the \ce{Sm2Co17} cell size \textit{L}, thickness of the stripe-shaped \ce{SmCo5} phase \textit{w}$_s$, the distance between Z-platelets \textit{d}, thickness of the Z-platelets \textit{w}$_z$, and orientation angle $\alpha$ between the field $\H$ and the easy axis $\eau$, as shown in \figref{fig3}a. In this work, we take $L=150~\si{nm}$, $d=50~\si{nm}$, $w_\mathrm{s}=w_\mathrm{Z}=8~\mathrm{nm}$, while $\alpha$ varies between 0 and $\pi/2$. The nanostructure is generated in a $512\times512\times4~\si{nm^3}$ finite difference domain. The periodic boundary condition (BC) is applied on the two boundaries perpendicular to the z direction, while the Neumann BC is applied on other boundaries. A grain boundary layer with the thickness of 2  \si{nm} where magnetocrystalline isotropy is assumed (i.e., $\Ku$), is also introduced to emulate the effects of the grain boundary in reducing the nucleation field to the system \cite{Yi2016a}. To recapture the domain wall behaviors in the micromagnetic simulations without artificial effects related to mesh, the FD cell size is chosen as 0.8 \si{nm}, which is smaller than the minimum magnetocrystalline exchange length. Micromagnetic parameters of each phase are presented in Table. \ref{tab1}.

\begin{table}[ht]
\caption{Micromagnetic parameters for the phases appearing in this work.}
\begin{tabular}{lllll}\hline                                                                   & Unit      & \ce{Sm2Co17} & \ce{SmCo5} & Z-platetes \\
\hline
$\Ae$ & \si{pJ~m^{-1}}          & 19.6                          & 8.6                         & 0.7        \\
$\Ku$ & \si{MJ~m^{-3}}          & 3.9                           & 18.3                        & 1.4        \\
$M_\mathrm{sat}$ & \si{kA~m^{-1}} & 987.7                         & 810.8                       & 310.4      \\
$\lex$ & nm                     & 7.0                           & 2.2                         & 2.2        \\
$\sigdw$ & \si{mJ~m^{-2}}            & 35.0                          & 50.2                        & 4.0    \\
\hline
\end{tabular}
\label{tab1}
\end{table}

\figref{fig3}b presents a half-cycle hysteresis of a nanostructure with $\alpha=\pi/6$ examined over a single demagnetizing process ($H$ from positive to negative with direction denoted in \figref{fig3}a). The magnetization reversal of the nanostructure consists of two steps: the reversed domain is firstly generated (nucleated) when the magnetic field reaches a certain threshold, denoted as the nucleation contribution $\Hn$. Then, the nucleated reversed domain starts to grow alongside the reversely increasing magnetic field, demonstrated in the form of domain wall migration. When the migrated domain wall front encounters the intersections between different phases where the domain wall energy differences exist, magnetic energy is consumed to compensate such differences and the domain wall front stops migration, i.e., the domain wall pinning occurs. The pinning events are reflected on the hysteresis curve as multiple stages where the magnetization is barely changed, as shown in the \figref{fig3}b. Therefore, the extra magnetic field (denoted as pinning contribution $\Hp$) is required for the magnetization reversal. 

We further present that the nucleation and pinning events on the nanostructure vary with the orientation angle, even though the parameters of constituent phases do not change. As shown in \figref{fig3}c, the half-cycle hysteresis curves present varying staging patterns w.r.t. $\alpha$, resulting in $\Hsw$ as a function of $\alpha$ as shown in the inset of \figref{fig3}c. We then take this $\Hsw(\alpha)$ and feed in the vector hysteron in \eqref{eq:VH} and present its longitudinal magnetization for comparison. We can tell that the hysteron can nicely emulate the magnetization reversal for the coherent case ($\alpha=0$). For increasing $\alpha$ to $\pi/2$, the hysteron presents an increased deviation in demagnetization compared to the micromagnetic simulation. When $H<\Hsw$, the hysteron shows less longitudinal demagnetization than the micromagnetic one, implying the relatively slower rotation of the surrogate magnetization vector; when $H>\Hsw$, the hysteron shows higher longitudinal demagnetization than the micromagnetic one, implying the relatively faster rotation of the surrogate magnetization vector. This difference in the demagnetization process between the hysteron and the micromagnetics eventually leads to the deviation of the magnetic coercivity for $0<\alpha<\pi/2$. For $\alpha=\pi/2$, both result in the zero coercivity, even though the difference in demagnetization process still exists.

\subsection{Hysteresis of Sm-Co Polycrystal}

\begin{figure}[!t]
\centerline{\includegraphics[width=8cm]{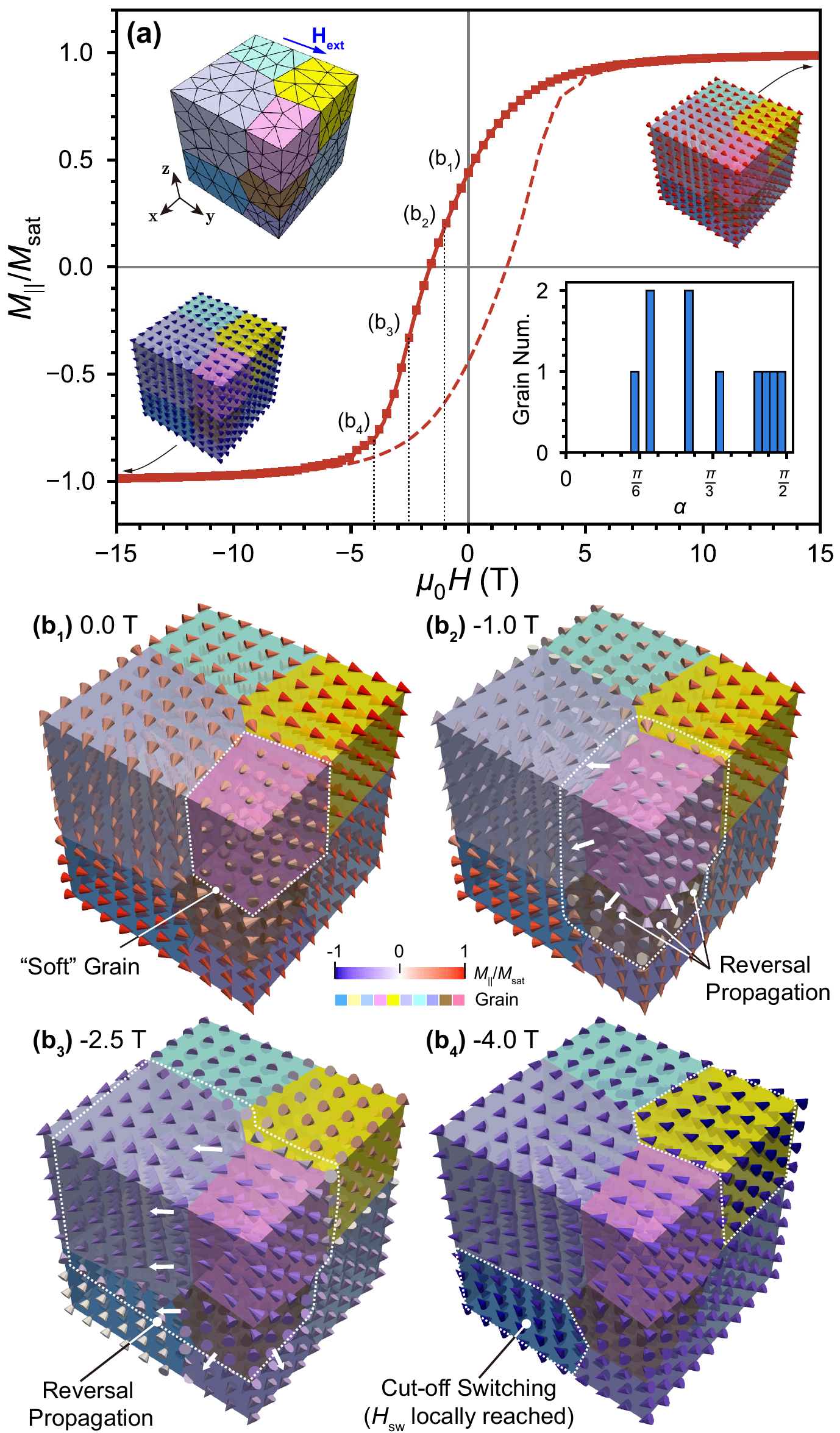}}
\caption{(a) The simulated hysteresis loop of the polycrystal structure. Insets: the FE mesh and the orientation histogram of the structure. Local magnetization reversal of sampled points (visualized as rotating cones) is also visualized under different applied field: (b$_1$) 0.0 T; (b$_2$) -1.0 T; (b$_3$) -2.5 T; (b$_4$) -4.0 T. }
\label{fig4}
\end{figure}

We apply the surrogate hysterons with micromagnetics-informed $\Hsw(\alpha)$ in a 10-grain polycrystal structure with the size $100\times100\times100~\si{\micro m^3}$. It is sufficiently large so that every point inside the polycrystal structure can be conceptually regard as the homogenized point of the local nanostructure. \figref{fig4}a presents the hysteresis loop of the structure with its mesh and orientation histogram shown in the insets. The magnetic coercivity of the polycrystal reads as 1.65 T, which is smaller than the one of 3.17 T from micromagnetic simulation on the coherent nanostructure ($\H\|\eau$). This is because more than half of the grains inside the examined polycrystal possess the orientation angles that are beyond $\pi/4$, which significantly affect the overall hystersis of the structure.

To deliver insight into how those grains with relatively larger $\alpha$ affect the demagnetization of the structure, we sample $10\times10\times10$ points  and visualize their on-site hysterons as oriented cones. With the applied magnetic field reduced to zero, we can tell that the local hysterons inside certain grains (denoted in \figref{fig4}b$_1$) point to the direction almost $\pi/2$ w.r.t. the magnetic field direction, which is also the easy axis of the grain. This grain is regarded as the ``soft'' grain as the local coercivity  inside is nearly zero. The surrounding hysterons receive the influence of the grain and deviate slightly from their stable directions. It is worth noting that the hysterons inside the grain with $\pi/4<\alpha<\pi/2$ would be affected relatively easier, presenting a trend of reversal propagation towards those directions.
 When the field starts to reversely increase ($H<0$), the hysterons inside the ``soft'' grain already present the reversed magnetization and continue propagating the effect to the surrounded grains with relatively large $\alpha$, as shown from \figref{fig4}b$_2$ to \ref{fig4}b$_3$. Meanwhile, the hysterons inside grains with $0\leq \alpha<\pi/4$ (``hard'' grains) receive less effects from already reversed ones, until the local field is large enough to suddenly reverse all of them, as denoted in \figref{fig4}b$_4$. This is due to the cut-off switching of those hysterons as demonstrated in \figref{fig2}, where the local field reaches the $\Hsw$.

\section{Conclusions}
We present in this work a novel multiscale simulation scheme for permanent magnets recapitulating its structural and physical characteristics from the nanoscale to the mesoscale. We perform the micromagnetic simulations on the parameterized Sm-Co nanostructures and investigate the mechanisms that are tightly related to the local magnetic coercivity, including the nucleation of the reversed domain, domain wall migration and pinning and unveil the orientation dependence of the demagnetization processes via half-cycle hysteresis curves. Those information are then carried by micromagnetics-informed surrogate hysterons in the magnetostatic homogenization of a 10-grain polycrystalline structure with assigend orientations. The simulation results of the polycrystal demonstrate that the grains with largest $\alpha$ (``soft'' grains) influence  the overall demagnetization process significantly by their early reversing and further propagate such effect to the grains with $\pi/4<\alpha<\pi/2$ via affecting the local field. This is believed to have the resulting magnetic coercivity of $1.65~\si{T}$ which is smaller than the one of $3.17~\si{T}$ examined from coherent nanostructure.

The presented work also depicts the demand of integrating data-driven methods, as the parameters of the surrogate hysteron intrinsically depend on the nanostructure and its orientation. In other words, effects of the nanostructural parameters (here the $L$, $d$, $w_\mathrm{s}$ and $w_\mathrm{Z}$) on the behavior of surrogate hysterons and the sensitivity analysis should be addressed in the upcoming works. More hysteron parameters that further help the surrogate hysteron to emulate the micromagnetic-simulated magnetization reversal, e.g., an orientation-dependent offset that adjusts the demagnetization (or rotation) of the hysteron, should be also discussed and examined.

\section*{Acknowledgements}
Authors acknowledge the financial support of German Science Foundation (DFG) in the framework of the Collaborative Research Centre Transregio 270 (CRC-TRR 270, project number 405553726, sub-projects A06) and 361 (CRC-TRR 361, project number 492661287, sub-projects A05). The authors also greatly appreciate their access to the Lichtenberg High-Performance Computer and the technique supports from the HHLR, Technische Universit\"at Darmstadt, and the GPU Cluster from the CRC-TRR 270 sub-project Z-INF.

\bibliographystyle{ieeetran}

\begin{thebibliography}{10}
	\providecommand{\url}[1]{#1}
	\csname url@samestyle\endcsname
	\providecommand{\newblock}{\relax}
	\providecommand{\bibinfo}[2]{#2}
	\providecommand{\BIBentrySTDinterwordspacing}{\spaceskip=0pt\relax}
	\providecommand{\BIBentryALTinterwordstretchfactor}{4}
	\providecommand{\BIBentryALTinterwordspacing}{\spaceskip=\fontdimen2\font plus
		\BIBentryALTinterwordstretchfactor\fontdimen3\font minus
		\fontdimen4\font\relax}
	\providecommand{\BIBforeignlanguage}[2]{{%
			\expandafter\ifx\csname l@#1\endcsname\relax
			\typeout{** WARNING: IEEEtran.bst: No hyphenation pattern has been}%
			\typeout{** loaded for the language `#1'. Using the pattern for}%
			\typeout{** the default language instead.}%
			\else
			\language=\csname l@#1\endcsname
			\fi
			#2}}
	\providecommand{\BIBdecl}{\relax}
	\BIBdecl
	
	\bibitem{duerrschnabel_atomic_2017}
	M.~Duerrschnabel, M.~Yi, K.~Uestuener, M.~Liesegang, M.~Katter, H.-J. Kleebe,
	B.~Xu, O.~Gutfleisch, and L.~Molina-Luna, ``Atomic structure and domain wall
	pinning in samarium-cobalt-based permanent magnets,'' \emph{Nature
		Communications}, vol.~8, no.~1, p.~54.
	
	\bibitem{katter1996new}
	M.~Katter, J.~Weber, W.~Assmus, P.~Schrey, and W.~Rodewald, ``A new model for
	the coercivity mechanism of sm/sub 2/(co, fe, cu, zr)/sub 17/magnets,''
	\emph{IEEE Transactions on Magnetics}, vol.~32, no.~5, pp. 4815--4817, 1996.
	
	\bibitem{song_cell-boundary-structure_2020}
	X.~Song, Y.~Liu, A.~Xiao, T.~Yuan, and T.~Ma,
	``\BIBforeignlanguage{en}{Cell-boundary-structure controlled
		magnetic-domain-wall-pinning in 2:17-type {Sm}-{Co}-{Fe}-{Cu}-{Zr} permanent
		magnets},'' \emph{\BIBforeignlanguage{en}{Materials Characterization}}, vol.
	169, p. 110575, Nov. 2020.
	
	\bibitem{wang_overview_2021}
	C.~Wang and M.-G. Zhu, ``Overview of composition and technique process study on
	2:17-type {Sm}–{Co} high-temperature permanent magnet,'' \emph{Rare
		Metals}, vol.~40, no.~4, pp. 790--798, Apr. 2021.
	
	\bibitem{zhou_revisiting_2021}
	X.~Zhou, Y.~Liu, W.~Jia, X.~Song, A.~Xiao, T.~Yuan, F.~Wang, J.~Fan, and T.~Ma,
	``Revisiting the pinning sites in 2:17-type sm-co-fe-cu-zr permanent
	magnets,'' \emph{Journal of Rare Earths}, vol.~39, no.~12, pp. 1560--1566.
	
	\bibitem{giro2022}
	S.~Giron, N.~Polin, E.~Adabifiroozjaei, Y.~Yang, D.~Ohmer, R.~U., K.~Üstüner,
	M.~Katter, I.~A. Radulov, K.~P. Skokov, B.-X. Xu, L.~Molina-Luna, B.~Gault,
	and O.~Gutfleisch, ``Investigation of two-step demagnetisation behavior of
	samarium-cobalt based sintered permanent magnets with high remanence,''
	\emph{(preparing)}, 2022.
	
	\bibitem{tuprints22038}
	\BIBentryALTinterwordspacing
	Q.~Gong, ``\BIBforeignlanguage{en}{Multiscale calculations of intrinsic and
		extrinsic properties of permanent magnets},'' Ph.D. dissertation, Technische
	Universit{\"a}t Darmstadt, Darmstadt, 2022. [Online]. Available:
	\url{http://tuprints.ulb.tu-darmstadt.de/22038/}
	\BIBentrySTDinterwordspacing
	
	\bibitem{preisach1935magnetische}
	F.~Preisach, ``{\"U}ber die magnetische nachwirkung,'' \emph{Zeitschrift
		f{\"u}r physik}, vol.~94, no.~5, pp. 277--302, 1935.
	
	\bibitem{takacs2001phenomenological}
	J.~Tak{\'a}cs, ``A phenomenological mathematical model of hysteresis,''
	\emph{COMPEL-The international journal for computation and mathematics in
		electrical and electronic engineering}, 2001.
	
	\bibitem{jiles_theory_1984}
	D.~C. Jiles and D.~L. Atherton, ``Theory of ferromagnetic hysteresis
	(invited),'' vol.~55, no.~6, pp. 2115--2120.
	
	\bibitem{zirka_physical_2012}
	S.~E. Zirka, Y.~I. Moroz, R.~G. Harrison, and K.~Chwastek, ``On physical
	aspects of the jiles-atherton hysteresis models,'' vol. 112, no.~4, p.
	043916.
	
	\bibitem{kronmuller2003micromagnetism}
	H.~Kronmuller, H.~Kronm{\"u}ller \emph{et~al.}, \emph{Micromagnetism and the
		microstructure of ferromagnetic solids}.\hskip 1em plus 0.5em minus
	0.4em\relax Cambridge university press, 2003.
	
	\bibitem{brown1963thermal}
	W.~F. Brown~Jr, ``Thermal fluctuations of a single-domain particle,''
	\emph{Physical review}, vol. 130, no.~5, p. 1677, 1963.
	
	\bibitem{sun2000}
	J.~Z. Sun, ``Spin-current interaction with a monodomain magnetic body: A model
	study,'' \emph{Physical Review B}, vol.~62, no.~1, pp. 570--578.
	
	\bibitem{leliaert2018fast}
	J.~Leliaert, M.~Dvornik, J.~Mulkers, J.~De~Clercq, M.~Milo{\v{s}}evi{\'c}, and
	B.~Van~Waeyenberge, ``Fast micromagnetic simulations on gpu—recent advances
	made with,'' \emph{Journal of Physics D: Applied Physics}, vol.~51, no.~12,
	p. 123002, 2018.
	
	\bibitem{goldfarb2009curvilinear}
	D.~Goldfarb, Z.~Wen, and W.~Yin, ``A curvilinear search method for p-harmonic
	flows on spheres,'' \emph{SIAM Journal on Imaging Sciences}, vol.~2, no.~1,
	pp. 84--109, 2009.
	
	\bibitem{Vansteenkiste2014}
	A.~Vansteenkiste, J.~Leliaert, M.~Dvornik, M.~Helsen, F.~Garcia-Sanchez, and
	B.~Van~Waeyenberge, ``The design and verification of mumax3,'' \emph{AIP
		Advances}, vol.~4, no.~10.
	
	\bibitem{exl2014labonte}
	L.~Exl, S.~Bance, F.~Reichel, T.~Schrefl, H.~Peter~Stimming, and N.~J. Mauser,
	``Labonte's method revisited: An effective steepest descent method for
	micromagnetic energy minimization,'' \emph{Journal of Applied Physics}, vol.
	115, no.~17, p. 17D118, 2014.
	
	\bibitem{Yi2016a}
	M.~Yi, O.~Gutfleisch, and B.~X. Xu, ``Micromagnetic simulations on the grain
	shape effect in nd-fe-b magnets,'' \emph{Journal of Applied Physics}, vol.
	120, no.~3.
	
	\bibitem{petrila2011analytical}
	I.~Petrila and A.~Stancu, ``Analytical vector generalization of the classical
	stoner--wohlfarth hysteron,'' \emph{Journal of Physics: Condensed Matter},
	vol.~23, no.~7, p. 076002, 2011.
	
	\bibitem{tonks2012object}
	M.~R. Tonks, D.~Gaston, P.~C. Millett, D.~Andrs, and P.~Talbot, ``An
	object-oriented finite element framework for multiphysics phase field
	simulations,'' \emph{Comput. Mater. Sci.}, vol.~51, no.~1, pp. 20--29, 2012.
	
	\bibitem{permann2020moose}
	C.~J. Permann, D.~R. Gaston, D.~Andr{\v{s}}, R.~W. Carlsen, F.~Kong, A.~D.
	Lindsay, J.~M. Miller, J.~W. Peterson, A.~E. Slaughter, R.~H. Stogner, and
	R.~C. Martineau, ``{MOOSE}: Enabling massively parallel multiphysics
	simulation,'' \emph{{SoftwareX}}, vol.~11, p. 100430, 2020.
	
	\bibitem{quey2011large}
	R.~Quey, P.~Dawson, and F.~Barbe, ``Large-scale 3d random polycrystals for the
	finite element method: Generation, meshing and remeshing,'' \emph{Computer
		Methods in Applied Mechanics and Engineering}, vol. 200, no. 17-20, pp.
	1729--1745, 2011.
	
	\bibitem{quey:hal-01626440}
	R.~Quey and L.~Renversade, ``{Optimal polyhedral description of 3D
		polycrystals: method and application to statistical and synchrotron X-ray
		diffraction data},'' \emph{{Computer Methods in Applied Mechanics and
			Engineering}}, vol. 330, pp. 308--333, Oct. 2017.
	
	\bibitem{quey:hal-01850591}
	R.~Quey, A.~Villani, and C.~Maurice, ``{Nearly uniform sampling of crystal
		orientations},'' \emph{{Journal of Applied Crystallography}}, vol.~51, no.~4,
	pp. 1162 -- 1173, Aug. 2018.
	
\end{thebibliography}


\end{document}